\documentclass[10pt,twocolumn,a4paper]{article}

\usepackage{times}
\usepackage{geometry}
\usepackage{authblk} 
\usepackage{amsmath,amssymb,amsthm}

\usepackage{graphicx}
\graphicspath{{figures/}}
\usepackage{multirow}
\usepackage{booktabs}

\usepackage{algorithm}
\usepackage{algorithmic}

\usepackage{cite}

\usepackage[hidelinks]{hyperref}

\title{\textbf{On Systematic Performance of 3-D Holographic MIMO: Clarke, Kronecker, and 3GPP Models}}
\author[1]{Quan Gao}
\author[2]{Shuai S. A. Yuan}
\author[1]{Zhanwen Wang}
\author[3]{Wanchen Yang}
\author[1]{Chongwen Huang}
\author[4]{Xiaoming Chen}
\author[1]{Wei E. I. Sha}

\affil[1]{College of Information Science and Electronic Engineering, Zhejiang University, Hangzhou 310027, China}
\affil[2]{Department of Electronics and Nano Engineering, School of Electrical Engineering, Aalto University, 02150 Espoo, Finland}
\affil[3]{College of Electronic and Information Engineering, Nanjing University of Aeronautics and Astronautics, Nanjing 211106, China}
\affil[4]{School of Information and Communication Engineering, Xi’an Jiaotong University, Xi’an 710049, China}

\date{}

\begin{document}
\twocolumn[
\maketitle
\vspace{-1em}
\begin{abstract}
Holographic multiple-input multiple-output (MIMO) has emerged as a key enabler for 6G networks, yet conventional planar implementations suffer from spatial correlation and mutual coupling at sub-wavelength spacing, which fundamentally limit the effective degrees of freedom (EDOF) and channel capacity. Three-dimensional (3-D) holographic MIMO offers a pathway to overcome these constraints by exploiting volumetric array configurations that enlarge the effective aperture and unlock additional spatial modes. This work presents the first systematic evaluation that jointly incorporates electromagnetic (EM) characteristics, such as mutual coupling and radiation efficiency, into the analysis of 3-D arrays under Clarke, Kronecker, and standardized 3rd Generation Partnership Project (3GPP) channel models. Analytical derivations and full-wave simulations demonstrate that 3-D architectures achieve higher EDOF, narrower beamwidths, and notable capacity improvements compared with planar baselines. In 3GPP urban macro channels with horizontal element spacing of $0.3\lambda$, 3-D configurations yield approximately 20\% capacity improvement over conventional 2-D arrays under the same element spacing and aperture size, confirming the robustness and scalability of volumetric designs under realistic conditions. These findings bridge the gap between theoretical feasibility and practical deployment, offering design guidance for next-generation 6G base station arrays.
\end{abstract}
\vspace{1em}
]
\maketitle

\section{Introduction}
Multiple-input multiple-output (MIMO) technology has become a cornerstone of modern wireless communications, significantly enhancing channel capacity through spatial multiplexing~\cite{telatar1999capacity,tse2005fundamentals}. Building on this foundation, the concept of massive MIMO was introduced, equipping base stations with hundreds of antennas to achieve highly directional beamforming and support large-scale multiuser connectivity~\cite{larsson2014massive,bjornson2019massive}. Massive MIMO has been widely recognized as a key enabling technology for 5G, offering substantial gains in spectral efficiency and energy efficiency~\cite{ngo2013energy}. However, as the wireless community moves toward 6G, the demand for even higher capacity, lower latency, and more flexible wave manipulation continues to grow~\cite{chowdhury20206g}. 

In response to these evolving demands, holographic MIMO has recently been proposed as an advanced architecture that integrates a massive number of tightly spaced, sub-wavelength antenna elements within a compact aperture~\cite{pizzo2020spatially,demir2022channel,wei2022multi,gong2024holographic,gong2024near}. This configuration enables highly fine-grained control over the electromagnetic (EM) field and offers enhanced flexibility in shaping wavefronts~\cite{huang2020holographic,zhang2023capacity}. Despite this potential, holographic MIMO faces several fundamental limitations. As antenna elements are placed at sub-wavelength spacing, strong spatial correlation arises, which reduces the effective degrees of freedom (EDOF)~\cite{pizzo2020degrees}. Moreover, the enhanced mutual coupling between adjacent elements leads to additional power loss and reduced radiation efficiency, ultimately constraining the achievable array gain. Given these fundamental constraints and considering that array apertures are often fixed in practical base station deployments, maximizing the performance of holographic MIMO within confined physical spaces has emerged as a critical research challenge.

A variety of approaches have been proposed to enhance holographic MIMO performance within confined apertures. One prominent direction leverages reconfigurable intelligent surfaces (RIS) to externally shape the wireless environment, thereby enhancing channel degrees of freedom~\cite{zhang2021large,cheng2022ris,su2023capacity}. However, RIS-based solutions require additional infrastructure and synchronized control between the base station and reflecting surfaces. Another research thread focuses on antenna element design, where decoupling techniques and metamaterial structures are employed to mitigate mutual coupling and improve radiation efficiency~\cite{jiang2023broadband,zhang2023multiband,zhou2024holographic,zhang2025multifunctional}. Recent work has explored advanced feeding structures and metasurface-based hybrid beamforming to enhance array gain while enabling efficient holographic wavefront control~\cite{shlezinger2019dynamic,an2023stacked,shao2025dual}. Beyond conventional radiators, novel antenna concepts have also been investigated, including fluid antennas that utilize liquid metals for dynamic aperture reconfiguration~\cite{new2024tutorial}, pinching antennas with mechanically tunable properties~\cite{yang2025pinching}, and Rydberg-atom-based arrays that offer unique electromagnetic characteristics~\cite{yuan2025electromagnetic}. While these element-level and environment-shaping approaches show promise, they often introduce additional hardware complexity, power consumption, or deployment constraints. In contrast, array topology optimization through three-dimensional element arrangement offers a more direct pathway: by extending deployments from planar to volumetric configurations, 3-D holographic MIMO can exploit oblique incidence to effectively enlarge the effective aperture, thereby increasing EDOF and channel capacity without requiring additional radio-frequency components or external control systems~\cite{yuan2024breaking, li2025capacity}.  

However, existing studies on 3-D holographic MIMO remain constrained by oversimplified assumptions. Many analyses adopt idealized isotropic antennas, neglecting critical EM characteristics such as mutual coupling and direction-dependent radiation patterns~\cite{costa2023ga,singh20243d}. Other works focus only on simplified propagation environments, such as line-of-sight (LoS) or spatially correlated Rayleigh fading channels~\cite{song2015spatial,yuan2023effects,yuan2025normalization}, which limit the generality of their conclusions. More fundamentally, no prior work has systematically evaluated 3-D holographic MIMO across the full spectrum from idealized theoretical models to standardized 3GPP propagation scenarios while simultaneously incorporating realistic antenna electromagnetic behavior. This gap hinders the translation of theoretical insights into deployable systems~\cite{gong2024jsac}.

To bridge this gap, this work develops a comprehensive framework for analyzing 3-D holographic MIMO arrays across progressively realistic channel models. In particular, we incorporate practical 3-D antenna designs and emphasize their electromagnetic characteristics, including mutual coupling and element radiation patterns, into the evaluation. The analysis begins with Clarke’s isotropic scattering model, proceeds to the Kronecker correlation model that captures spatial correlation and coupling effects, and finally extends to standardized 3GPP urban macro (UMa) scenarios that account for environment-specific parameters. Through these models, EDOF, channel capacity, and beamforming behavior are jointly examined, demonstrating how volumetric array configurations extend the performance envelope beyond planar counterparts.

The remainder of this paper is organized as follows. Section II introduces the fundamental principles of 3-D holographic MIMO, emphasizing the concepts of degrees of freedom and effective aperture enlargement. Section III develops a theoretical framework under three representative channel models (Clarke, Kronecker, and 3GPP) to progressively capture different levels of physical realism. Section IV incorporates practical antenna designs and their electromagnetic characteristics, including mutual coupling and radiation efficiency, into full-wave simulations for performance validation under realistic scenarios. Finally, Section V concludes the paper.

\section{Principles of 3-D Holographic MIMO}
In MIMO systems, a fundamental concept is the degree of freedom (DOF), which characterizes the number of parallel spatial channels that can be exploited for data transmission. According to \textit{Fundamentals of Wireless Communications}~\cite{tse2005fundamentals}, the DOF is mathematically defined as the rank of the channel correlation matrix
\begin{equation}
    \mathbf{\Psi} = \mathbf{H}\mathbf{H}^\dagger
\end{equation}
where $^\dagger$ denotes the conjugate transpose. The rank of $\mathbf{\Psi}$ is equivalent to the number of significant eigenvalues and reflects the maximum number of independent data streams that can be spatially multiplexed in a given MIMO system.

From a physical perspective, the DOF can be interpreted as the number of significant spatial EM modes supported by the communication environment. Radiation modes are inherently linked to the geometry and electrical size of the transmitting and receiving structures~\cite{gustafsson2024degrees}. Consequently, in the electrically large limit, the effective aperture seen in a given direction $\hat{k}$ approaches the geometric projection of the structure onto a plane normal to $\hat{k}$~\cite{gustafsson2019maximum}. For 3-D apertures, the DOF scales with the aperture area (in units of $\lambda^2$, where $\lambda$ is the free-space wavelength), whereas for 2-D (linear) apertures, it scales with length (in units of $\lambda$). A unified expression can thus be written as~\cite{gustafsson2025degrees}

\begin{equation}
    \mathrm{DOF}\;\approx\;
    \begin{cases}
        \dfrac{L_{\mathrm{eff}}}{\lambda}, & \text{2-D aperture} \\[2ex]
        \dfrac{A_{\mathrm{eff}}}{\lambda^2}, & \text{3-D aperture}
    \end{cases}
\end{equation}
where the effective length and aperture area are defined by integrating over all directions
\begin{equation}
    L_{\mathrm{eff}} = \int_{0}^{2\pi} \ell(\phi)\,\mathrm{d}\phi, 
    \qquad
    A_{\mathrm{eff}} = \int_{4\pi} \mathcal{P}(\hat{k}) \,\mathrm{d}\Omega_{\hat{k}}
\end{equation}
where $\ell(\phi)$ denotes the projection of the aperture length along the direction specified by $\phi$, $\mathcal{P}(\hat{k})$ denotes the projection of the aperture area along direction $\hat{k}$, and $\mathrm{d}\Omega_{\hat{k}}$ represents the differential solid angle.

\begin{figure}[ht!]
\centering
    \includegraphics[width=0.9\linewidth]{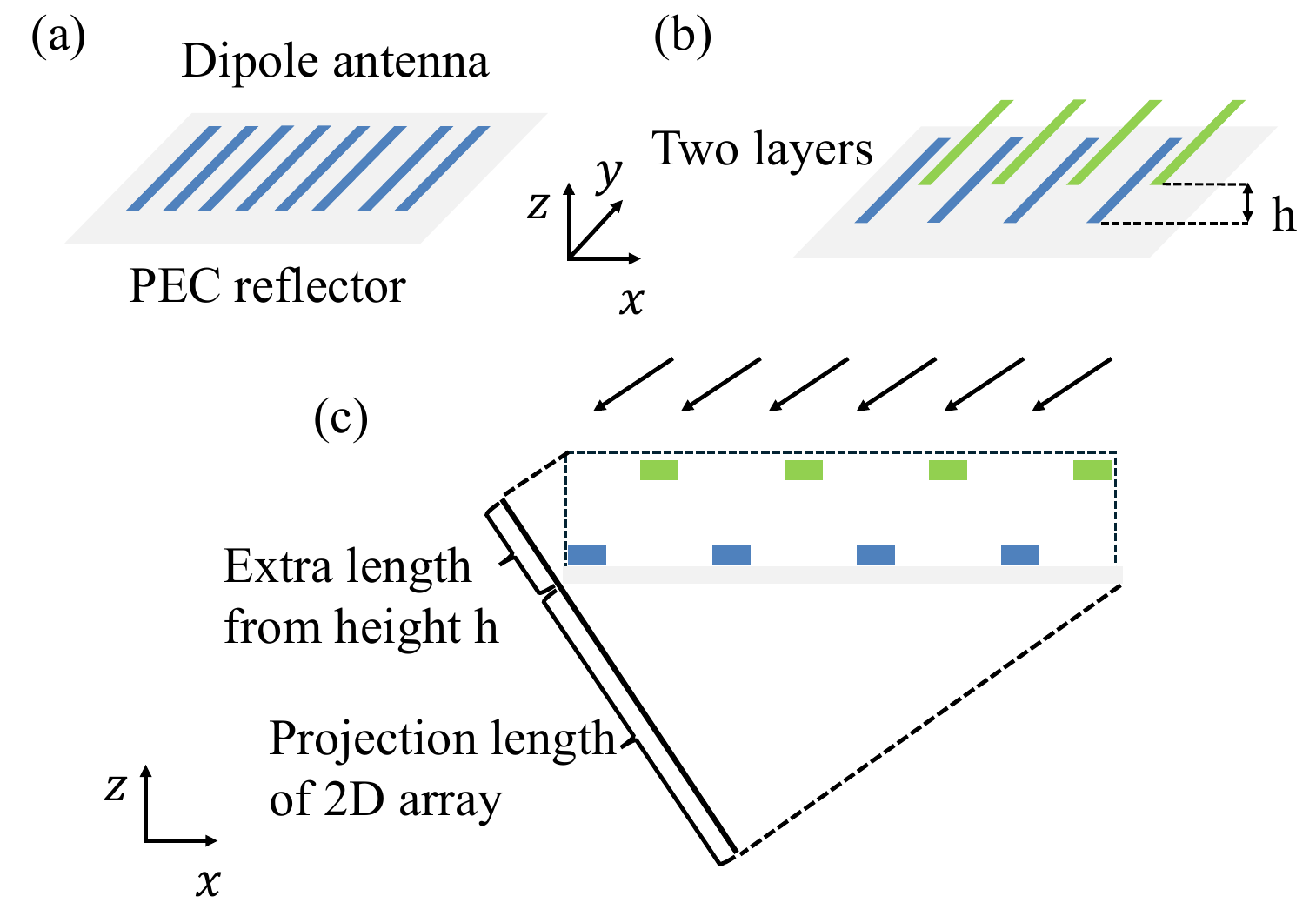}
\caption{Comparison between planar and volumetric linear arrays: (a) Single-layer dipole array above a PEC reflector, (b) Two-layer volumetric array with alternate element elevation, and (c) Effective projection enlargement under oblique incidence.}
\label{linear_array}
\end{figure}

\begin{figure}[ht!]
\centering
    \includegraphics[width=0.9\linewidth]{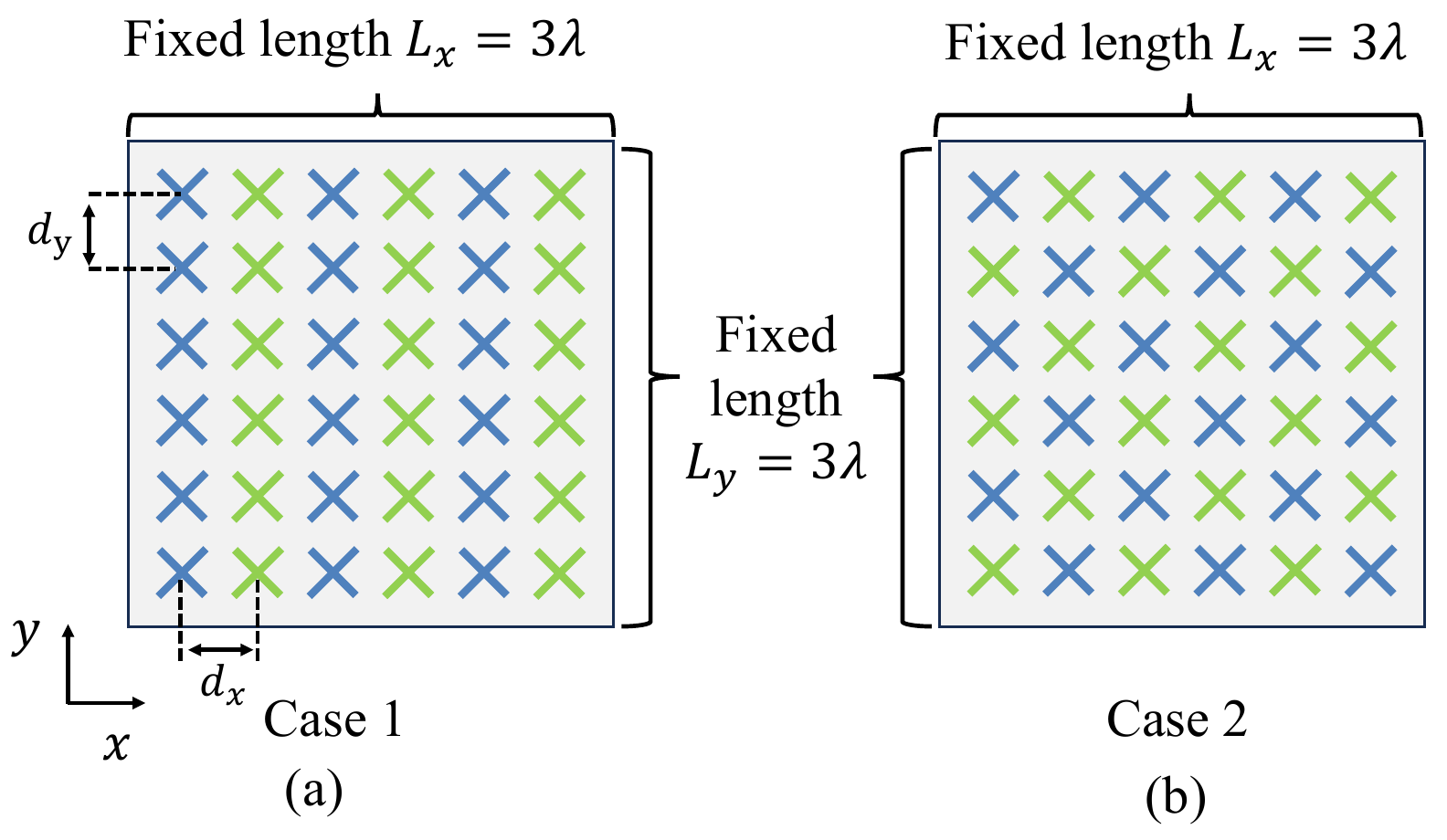}
\caption{Two 3-D antenna array configurations with fixed aperture $L_x =L_y=3\lambda$: (a) Case 1 with even-numbered columns elevated by height h, and (b) Case 2 with interleaved elevation pattern.}
\label{array_architecture}
\end{figure}

As an illustrative example, consider a single-row planar linear array in which the even-numbered elements are elevated by a height $h$ while the odd-numbered elements remain on the baseline, as shown in Figure~\ref{linear_array}. In this configuration, the staggered elevation introduces an additional effective projection area under oblique incidence [Figure~\ref{linear_array}(c)], which enhances the effective EDOF and consequently improves the channel capacity~\cite{yuan2024breaking}.

This volumetric design concept can be naturally extended to multi-row planar arrays in the $x$–$y$ plane, leading to two distinct 3-D configurations (see Figure~\ref{array_architecture}). 
In the first configuration (Case~1), all elements in even-numbered columns are elevated by $h$, while the odd-numbered columns remain on the baseline. 
In the second configuration (Case~2), an interleaved elevation is applied: elements in odd-numbered columns are raised only in odd-numbered rows, and those in even-numbered columns are elevated in even-numbered rows. 
Both 3-D structures reduce to the conventional planar array when $h=0$, which serves as the baseline for comparison in the subsequent analysis.

In addition to the DOF, the EDOF is frequently used to characterize spatial multiplexing performance. The EDOF can be interpreted as the equivalent number of independent single-input single-output (SISO) channels supported by the system. Mathematically, it is defined as
\begin{equation}
\Psi_e(\mathbf{\Psi}) = \left( \frac{\mathrm{tr}(\mathbf{\Psi})}{\|\mathbf{\Psi}\|_F} \right)^2 
= \frac{\left( \sum_i \sigma_i \right)^2}{\sum_i \sigma_i^2}
\end{equation}
where $\mathbf{\Psi} = \mathbf{H}\mathbf{H}^\dagger$ is the channel correlation matrix, $\mathrm{tr}(\cdot)$ denotes the trace operator, $\|\cdot\|_F$ is the Frobenius norm, and $\sigma_i$ are the eigenvalues of $\mathbf{\Psi}$.

Both DOF and EDOF can be used to assess the spatial multiplexing capability of MIMO systems. The DOF offers a more physically intuitive interpretation, as its value corresponds to the number of independent far-field electromagnetic modes that the array can support, which is primarily determined by the physical aperture size. By contrast, the EDOF provides a more practical system-level metric. It is highly sensitive to physical parameters such as antenna spacing; as spacing decreases, the resulting increase in spatial correlation leads to a more uneven eigenvalue distribution and a lower EDOF. From a communication perspective, EDOF represents the effective number of parallel data streams that can be efficiently supported. Although the number of RF chains often limits the number of concurrent streams in practice, the EDOF determines the quality and separability of these streams. A higher EDOF ensures that the system can fully utilize its available RF chains even in spatially constrained environments, directly leading to higher achievable capacity.

Therefore, compared with the geometrically defined DOF, the EDOF provides a more direct and practically relevant indicator for capacity analysis. In this context, the channel capacity can be approximated as~\cite{yuan2021electromagnetic}
\begin{equation}
C \approx B \Psi_e \log_2 \left( 1 + \frac{\gamma}{\Psi_e} \right)
\end{equation}
where $B$ denotes the system bandwidth and $\gamma$ is the total signal-to-noise ratio (SNR). This expression highlights that increasing EDOF does not automatically guarantee proportional capacity growth. Although 3-D configurations enlarge the set of accessible spatial modes and thereby increase the EDOF, mutual coupling and impedance mismatch may reduce embedded radiation efficiency and the effective SNR per spatial stream. Consequently, the achievable capacity is governed by a trade-off between the number of usable spatial modes and the power that can be efficiently delivered to each mode. This physical balance between EDOF enhancement and coupling-induced efficiency loss will be examined quantitatively in Sections~III and IV.

\section{Performance Analysis with Theoretical Antenna Models}
\begin{figure*}[t]
\centering
\includegraphics[width=0.9\textwidth]{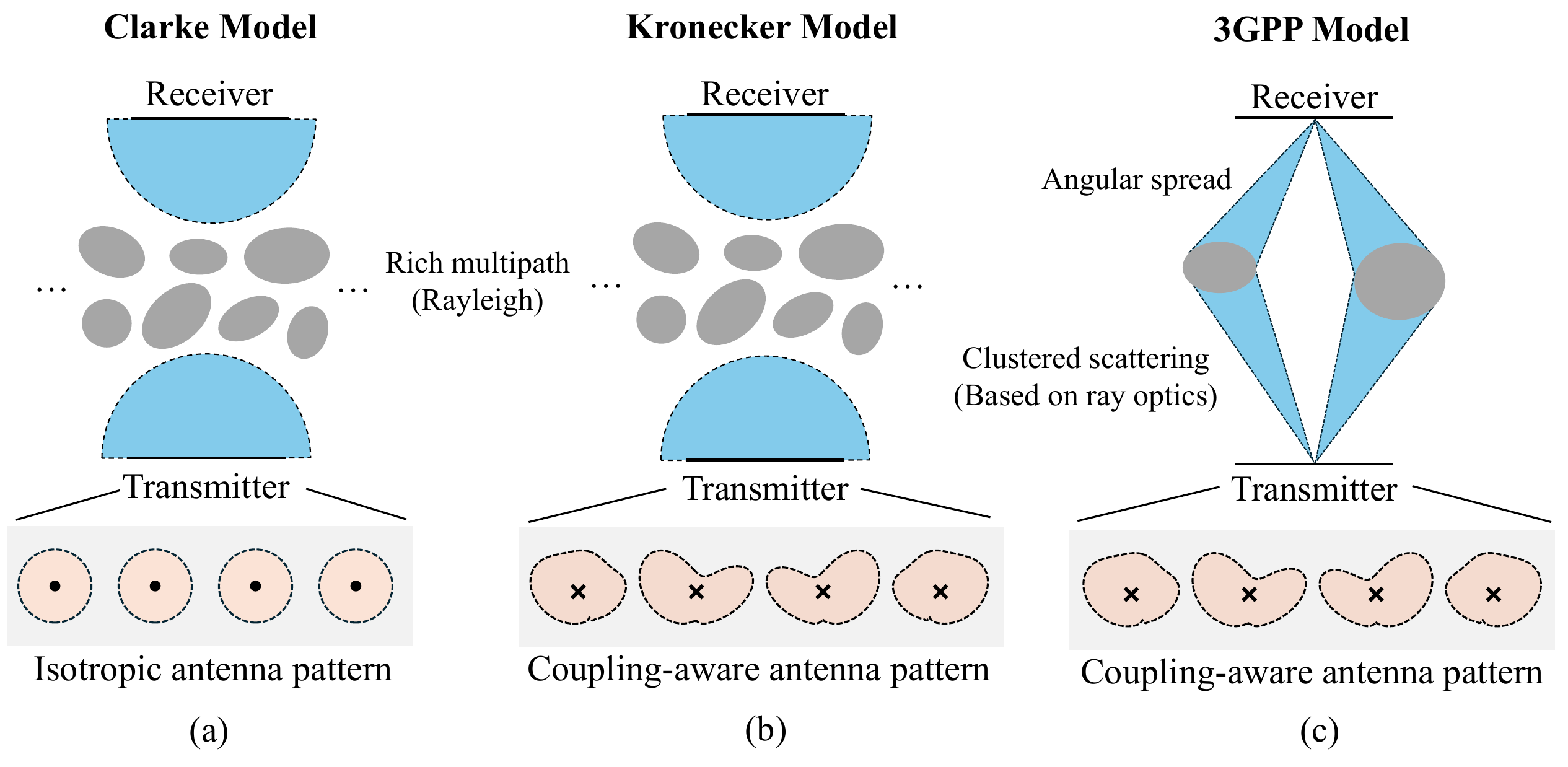}
\caption{Three representative channel models for MIMO performance analysis: (a) Clarke model with isotropic antenna patterns and rich Rayleigh multipath, (b) Kronecker model incorporating coupling-aware antenna responses, and (c) 3GPP model with clustered scattering based on ray optics.}
\label{clarke_kronecker_3gpp}
\end{figure*}

3-D holographic MIMO performance is evaluated using three channel models: Clarke, Kronecker, and 3GPP. 
These models are intentionally employed in a hierarchical manner, ranging from an idealized theoretical baseline to progressively more realistic antenna and propagation representations.
As illustrated in Figure~\ref{clarke_kronecker_3gpp}, the Clarke model provides the most fundamental description, where the wireless channel is constructed from rich multipath components generated by isotropic point scatterers, and both the transmit and receive antennas are assumed to be ideal and omnidirectional.drec 
Building upon this, the Kronecker model incorporates practical antenna characteristics by accounting for mutual coupling, radiation efficiency, and pattern distortion, thereby extending the Clarke model to more realistic array environments. 
The 3GPP model further advances the modeling framework by introducing clustered scattering based on ray optics and explicitly capturing angular spread, multipath fading, and path loss, which together reflect propagation conditions typically encountered in cellular networks.

\subsection{Clarke Model}
The Clarke model is one of the most widely used correlation models for wireless channels, providing a mathematical description of spatial correlation among antennas in multipath environments. In the 3-D case, antennas are modeled as point sources or receivers, and the impinging waves are modeled as uniformly distributed plane waves over the entire angular domain. Each received signal can thus be represented as the superposition of these plane waves arriving from different directions. Consequently, the spatial correlation between two antennas is fundamentally determined by their relative positions with respect to the incoming wavefronts.

Formally, the $(m,n)$-th entry of the correlation matrix $\mathbf{\Psi}$ quantifies the spatial correlation between the $m$-th and $n$-th antennas, and can be written as
\begin{equation}
    \rho_{nm} = \frac{1}{N} \int_{\Omega} \exp\!\left[j\mathbf{k}_{\Omega} \cdot (\mathbf{r}_{n} - \mathbf{r}_{m})\right] \,\mathrm{d}\Omega
\end{equation}
where $N$ is the number of plane waves, $\Omega$ denotes the solid angle of incident directions, and $\mathbf{k}_\Omega$ is the wave vector. In this formulation, only the multipath effect is considered. To capture thermal noise at the receiver, an additional additive white Gaussian noise (AWGN) matrix $\mathbf{H}_w$ can be applied after obtaining the correlation matrix. This expression reflects the fact that correlation decreases as the antenna separation increases, with the angular spread of the incoming multipath waves serving as the dominant factor.

Under the Clarke model, the ergodic capacity of a MIMO system can be evaluated based on the channel matrix statistics. Assuming rich scattering, Rayleigh fading, and equal power allocation across transmit antennas, we consider capacity normalized per unit bandwidth (i.e., spectral efficiency), and it is expressed as
\begin{equation}
\label{eq1}
    C = \mathbb{E}\left\{\log_2\left[\det\left(\mathbf{I} + \frac{\gamma}{N_t}\mathbf{H}\mathbf{H}^\dagger\right)\right]\right\}
\end{equation}
where $\mathbb{E}\{\cdot\}$ denotes expectation over channel realizations, $\mathbf{I}$ is the identity matrix, $\gamma$ is the total SNR, $N_t$ is the number of transmit antennas, and $\mathbf{H}$ represents the channel matrix.  

\begin{figure}[ht!]
\centering
    \includegraphics[width=0.9\linewidth]{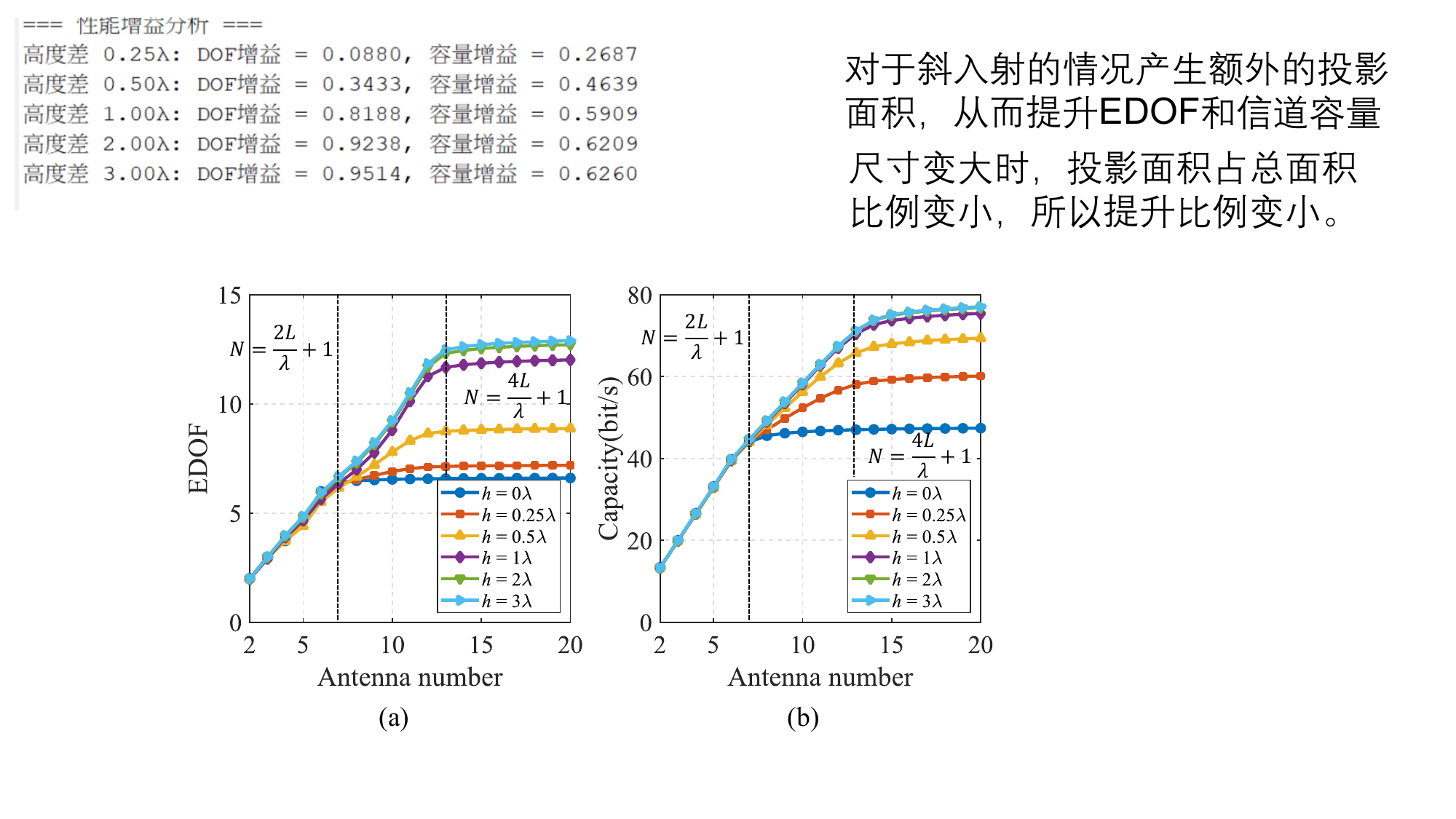}
\caption{Performance of linear arrays in the Clarke model under different vertical offsets $h$: (a) EDOF, (b) channel capacity.}
\label{clarke_linear}
\end{figure}

To gain insight into volumetric array behavior, a single-row linear array with fixed length $L = d \times N = 3\lambda$ is first analyzed, where $N$ is the number of elements. The SNR is set to 20 dB, while both the element spacing and the vertical offset $h$ are varied. As shown in Figure~\ref{clarke_linear}, the planar configuration ($h = 0$) saturates at the theoretical limit $N = \tfrac{2L}{\lambda} + 1$, corresponding to a horizontal projected spacing close to $0.5\lambda$, beyond which neither EDOF nor capacity increases. In contrast, volumetric arrays with nonzero elevation offsets continue to exhibit growth in both EDOF and capacity as the array is densified, eventually saturating at $N = \tfrac{4L}{\lambda} + 1$, where the horizontal projected spacing approaches $0.25\lambda$. 

\begin{figure}[ht!]
\centering
    \includegraphics[width=0.9\linewidth]{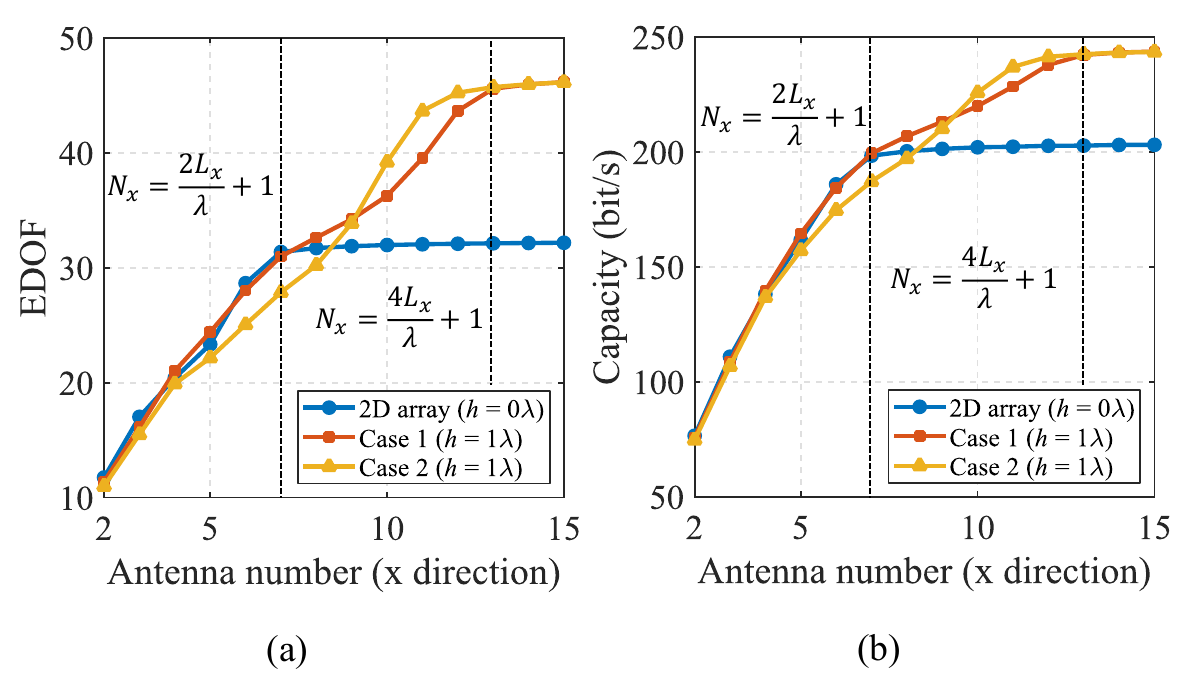}
\caption{Performance of 2-D planar and 3-D volumetric arrays in the Clarke model under densification along the $x$ direction: (a) EDOF, (b) channel capacity.}
\label{clarke_planar_oneside}
\end{figure}
\begin{figure}[ht!]
\centering
    \includegraphics[width=0.9\linewidth]{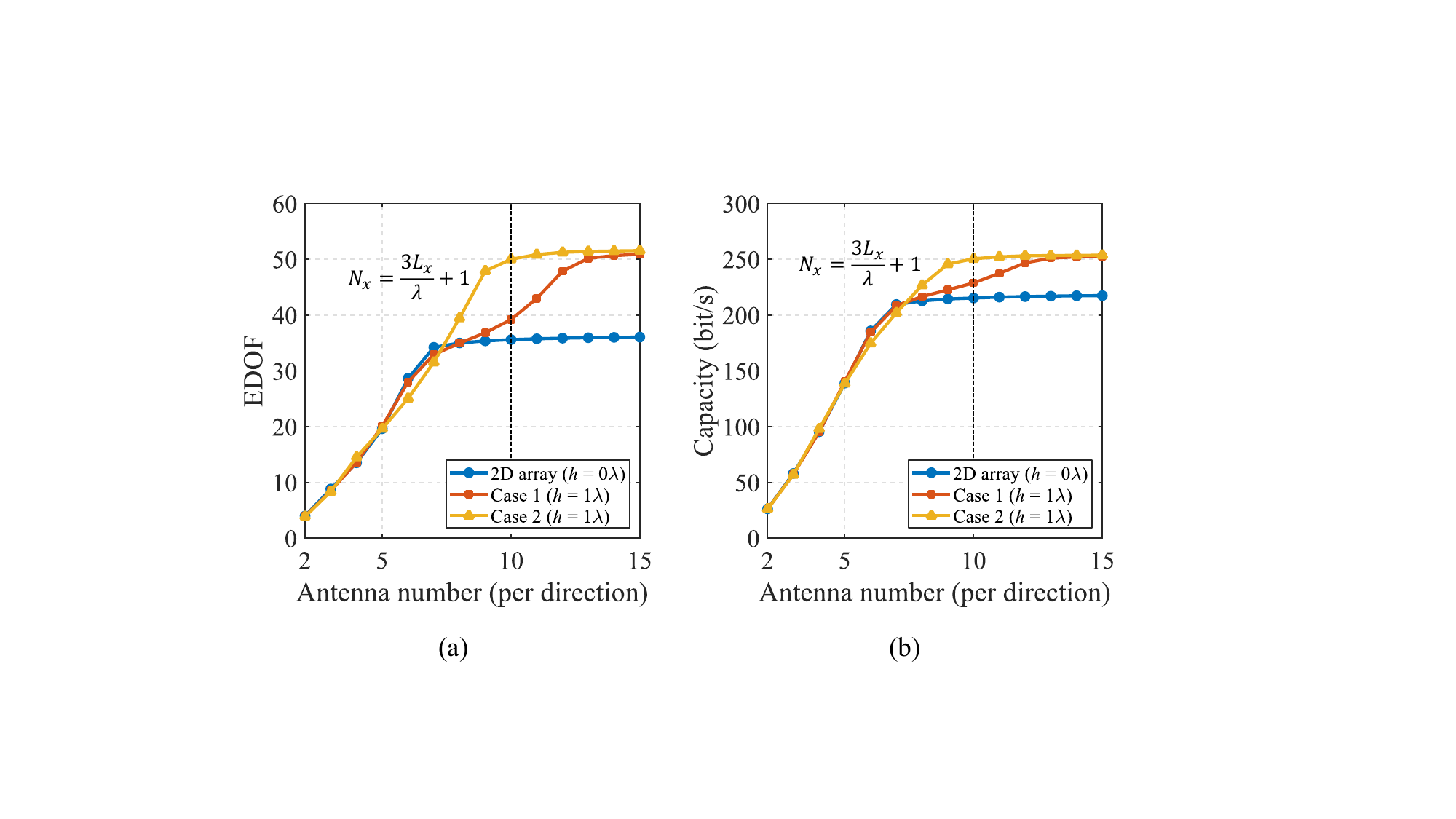}
\caption{Performance of 2-D planar and 3-D volumetric arrays in the Clarke model under simultaneous densification along both directions with $N_x = N_y$: (a) EDOF, (b) channel capacity.}
\label{clarke_planar_twoside}
\end{figure}

A quantitative comparison further highlights the benefits of introducing vertical offsets. When the offset is $0.25\lambda$, the EDOF and capacity limits increase by 8.88\% and 26.87\%, respectively. Increasing the offset to $0.5\lambda$ yields improvements of 34.33\% in EDOF and 46.39\% in capacity. With an offset of $1\lambda$, the enhancements reach 81.88\% and 59.09\%. However, further enlarging the vertical offset to $2\lambda$ or even $3\lambda$ produces only marginal additional gains. Therefore, in the subsequent analysis, the inter-layer vertical offset of 3-D arrays is fixed at $1\lambda$.

The analysis is subsequently extended to planar arrays with a fixed aperture of $S = L_x \times L_y = 3\lambda \times 3\lambda$. The element spacing in the $y$ direction is fixed at $0.5\lambda$, while the number of elements along the $x$ direction is gradually increased. The evaluation is performed with an elevation offset of $h = 1\lambda$ and an SNR of 20 dB. As illustrated in Figure~\ref{clarke_planar_oneside}, the overall trend is consistent with that observed in the linear array. A closer comparison indicates that Case~2 reaches its saturation point earlier than Case~1, highlighting a more efficient utilization of array elements. At saturation, both 3-D configurations yield approximately 44\% higher EDOF and 20\% higher capacity relative to the planar baseline.

Further densification in both the x and y directions while maintaining $N_x = N_y$ is presented in Figure~\ref{clarke_planar_twoside}. The results show that Case~2 reaches saturation at the same performance level much earlier than Case~1, specifically at $N = \frac{3L}{\lambda} + 1$. This behavior demonstrates that the advantage of Case~2 in reducing the number of required antennas becomes more evident when volumetric arrays are extended to densely packed 2-D configurations.

\subsection{Kronecker Model}
The Kronecker model provides a more generalized approximation for MIMO Rayleigh channels by assuming separable spatial correlations at the transmitter (Tx) and receiver (Rx). Under this assumption, the overall channel covariance matrix is expressed as the Kronecker product of the transmit and receive covariance matrices. The Vertical Bell Laboratories Layered Space-Time (V-BLAST) architecture~\cite{tse2005fundamentals} is commonly employed for Kronecker model analysis. It assumes ideal Tx antennas (uncorrelated with unit efficiency), i.e., $\mathbf{R}_t = \mathbf{I}_{N_t \times N_t}$, enabling focused evaluation of Rx array performance. Under this framework, the MIMO capacity is \cite{chen2015effect,chen2012mrc}:
\begin{equation}
\label{eq3}
    C = \mathbb{E}\left\{\log_2\left[\det\left(\mathbf{I} + \frac{\gamma}{N_t} \mathbf{R} \mathbf{H}_w \mathbf{H}_w^\dagger\right)\right]\right\}
\end{equation}
where the entries of $\mathbf{H}_w$ (matrix dimension is $N_t \times N_r$) are independent and identically distributed (i.i.d.) complex Gaussian variables denoting spatially white MIMO channel.

The Kronecker model incorporates practical antenna effects including pattern distortion from mutual coupling and radiation efficiency degradation. Mutual coupling modifies the port currents and therefore the embedded element radiation patterns as well as the per-port radiation efficiencies. Because these coupling effects are coherent and port-dependent, we extract embedded element patterns and embedded efficiencies from full-wave simulations and use them to form the channel covariance. To account for these effects, covariance matrix $\mathbf{R}$ is modified by the Hadamard (element-wise) product between the correlation matrix $\mathbf{\Psi}$ and the efficiency matrix $\mathbf{\Xi}$:
\begin{equation}
\label{eq4}
    \mathbf{R} = \mathbf{\Psi} \circ \mathbf{\Xi}
\end{equation}
where $\mathbf{\Psi}$ is the RX correlation matrix, and its $(m,n)$-th element $\rho_{nm}$ quantifies the correlation between the $n$-th and $m$-th antennas, which is expressed as~\cite{chen2017multiplexing}:
\begin{equation}
\label{eq5}
    \rho_{mn} = \frac{\oint G_{mn}(\Omega) \mathrm{d}\Omega}{\sqrt{\oint G_{mm}(\Omega) \mathrm{d}\Omega} \sqrt{\oint G_{nn}(\Omega) \mathrm{d}\Omega}}
\end{equation}
where 
\begin{equation}
\label{eq6}
    G_{mn}(\Omega) = \kappa E_{\theta m}(\Omega) E_{\theta n}^{*}(\Omega) P_{\theta}(\Omega) + E_{\phi m}(\Omega) E_{\phi n}^{*}(\Omega) P_{\phi}(\Omega)
\end{equation}
with $E_{\theta}$ and $E_{\phi}$ being the embedded element patterns for $\theta$/$\phi$ polarizations \cite{pozar2002active}, $P(\Omega)$ the angular power spectrum characterizing angle spread, and $\kappa$ the cross-polarization ratio.

The embedded efficiency matrix $\mathbf{\Xi}$ is defined as 
\begin{equation}
\label{eq7}
    \mathbf{\Xi} = \sqrt{\mathbf{e}} \sqrt{\mathbf{e}}^T
\end{equation}
where the efficiency vector
\begin{equation}
\label{eq8}
    \mathbf{e} = \left[e_{\text{emb}1}, e_{\text{emb}2}, \cdots, e_{\text{emb}N_{r}}\right]^{T}
\end{equation}
 When neglecting ohmic losses, the embedded radiation efficiency of the $n$-th antenna can be derived from $S$-parameters \cite{hannan2003element,kildal2015fundamental}:
\begin{equation}
\label{eq9}
    e_{\text{emb}n} = 1 - \sum_{i=1}^{N_r} |S_{in}|^2
\end{equation}

\begin{figure}[ht!]
\centering
    \includegraphics[width=0.8\linewidth]{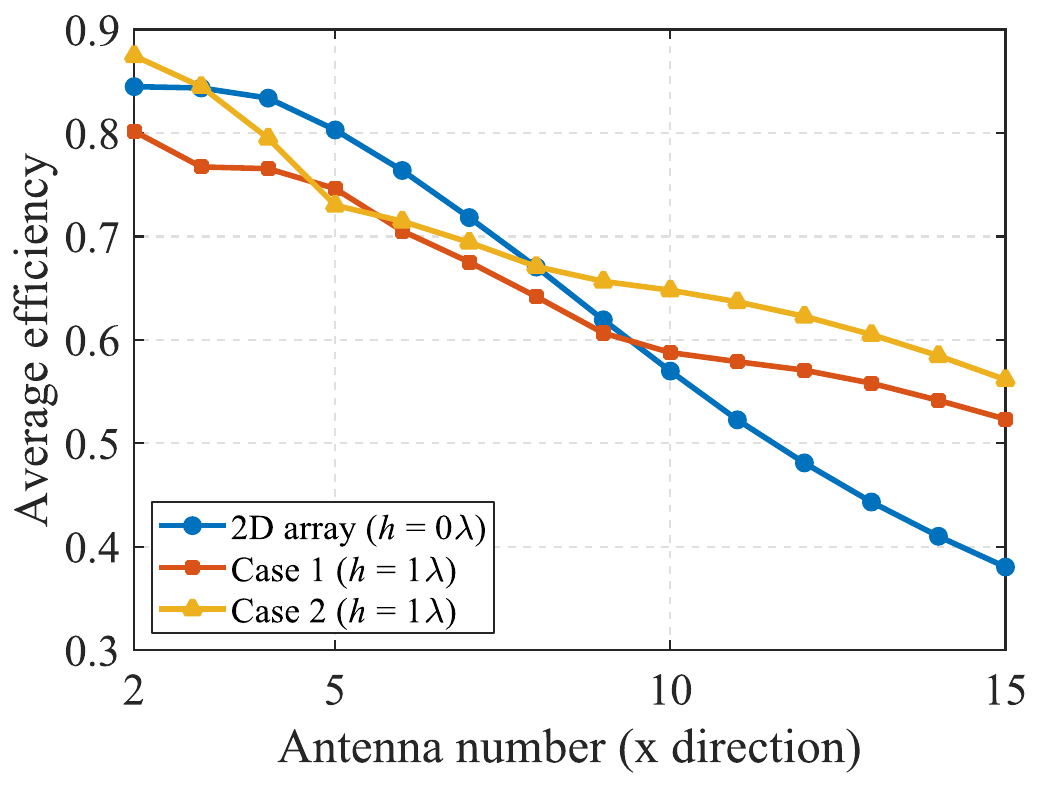}
\caption{Average radiation efficiency versus number of elements for different array configurations in the Kronecker model.}
\label{average_efficiency}
\end{figure}

\begin{figure}[ht!]
\centering
    \includegraphics[width=0.9\linewidth]{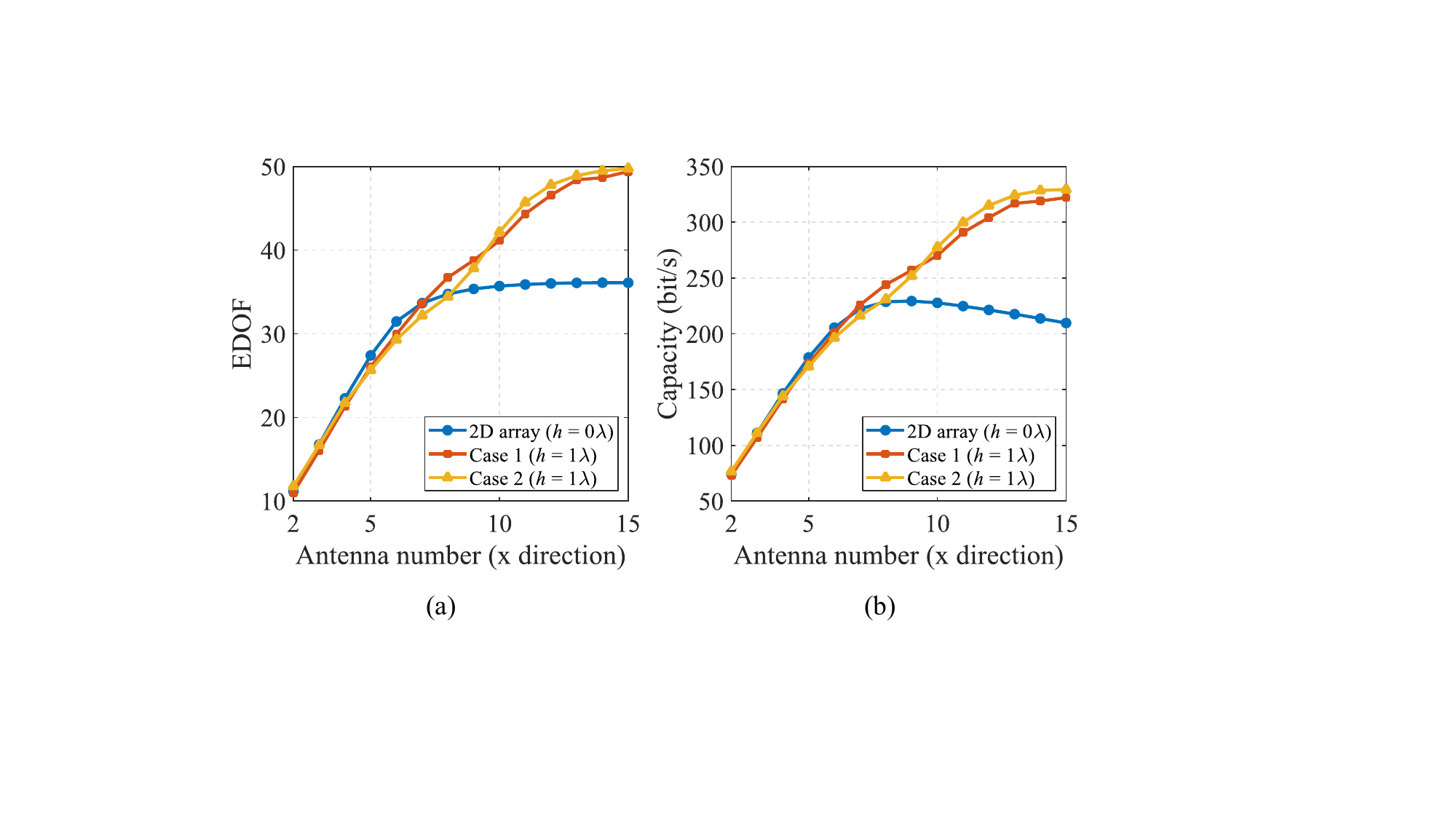}
\caption{Performance of 2-D planar and 3-D volumetric arrays in the Kronecker model under densification along the $x$ direction: (a) EDOF, (b) channel capacity.}
\label{kronecker_result}
\end{figure}

The Kronecker model simulation results are shown in Figures~\ref{average_efficiency} and~\ref{kronecker_result}. The array aperture, element spacing, and inter-layer height are identical; the only difference is that, unlike the Clarke model with ideal isotropic radiators, each element here is a dipole simulated in full-wave EM, capturing mutual coupling and radiation efficiency.

Figure~\ref{average_efficiency} illustrates average radiation efficiency versus the number of elements. As arrays densify, efficiency drops for all cases, with planar arrays degrading much faster than volumetric ones. The vertical separation in 3-D arrays redistributes coupling and alleviates the strongest interactions.

Figure~\ref{kronecker_result} presents the EDOF and channel capacity. The EDOF increases with densification and then saturates, similar to the Clarke model. However, since efficiency continues to fall, planar-array capacity peaks and then declines, while volumetric arrays maintain higher efficiency and delay this turning point. Thus, under strong coupling, 3-D arrays achieve larger capacity gains than predicted by the Clarke model.

\subsection{3GPP Model}
Under the 3GPP channel model, the achievable sum capacity of a $K$-user MU-MIMO system (with $K$ denoting the number of users) can be expressed as \cite{anselmi2022optimal}
\begin{equation}
\label{eq10}
    C_{\mathrm{MU-MIMO}}=\sum_{k=1}^{K}\operatorname{log}_{2}\left(1+\mathrm{SINR}_{k}\right)
\end{equation}
where $\mathrm{SINR}_{k}$ denotes the signal-to-interference-plus-noise ratio (SINR) of the $k$th user.
The SINR of the $k$th user is given by
\begin{equation}
\label{eq11}
\mathrm{SINR}_{k} = 
\frac{P_{\text{sig},k}}{P_{\text{int},k} + P_{\text{noise}}}
= \frac{\left\|\mathbf{h}_{k}\mathbf{w}_{k}\right\|^{2}}
{\sum_{j\neq k}\left\|\mathbf{h}_{k}\mathbf{w}_{j}\right\|^{2}+\sigma^{2}}
\end{equation}
where $P_{\text{sig},k}=\|\mathbf{h}_{k}\mathbf{w}_{k}\|^{2}$ denotes the received desired signal power for user $k$, 
$P_{\text{int},k}=\sum_{j\neq k}\|\mathbf{h}_{k}\mathbf{w}_{j}\|^{2}$ represents the multi-user interference (MUI) power caused by simultaneous transmissions to other users, and $P_{\text{noise}}=\sigma^{2}$ corresponds to the noise power. 
Here, $\mathbf{h}_{k}\in\mathbb{C}^{1\times U}$ is the channel vector between the base station with $U$ antennas and the $k$th user, 
$\mathbf{w}_{k}\in\mathbb{C}^{U\times 1}$ is the precoding vector assigned to user $k$, and $\sigma^2$ is the noise variance. 

The precoding matrix $\mathbf{W}$ is typically designed using Zero-Forcing (ZF) or Minimum Mean Square Error (MMSE) criteria. For the MMSE precoding, the matrix is given by
\begin{equation}
\label{eq12}
    \mathbf{W} = \mathbf{H}^\dagger \left( \mathbf{H} \mathbf{H}^\dagger + \alpha \mathbf{I} \right)^{-1}
\end{equation}
where $\alpha$ is a regularization factor. It is noted that in multi-user MIMO systems the performance gain of the 3-D arrays mainly arises from improved spatial separability among user channels, which enables the MMSE precoder to more effectively suppress inter-user interference, rather than from simple geometric beam steering toward individual users.

\begin{figure}[ht!]
\centering
    \includegraphics[width=0.9\linewidth]{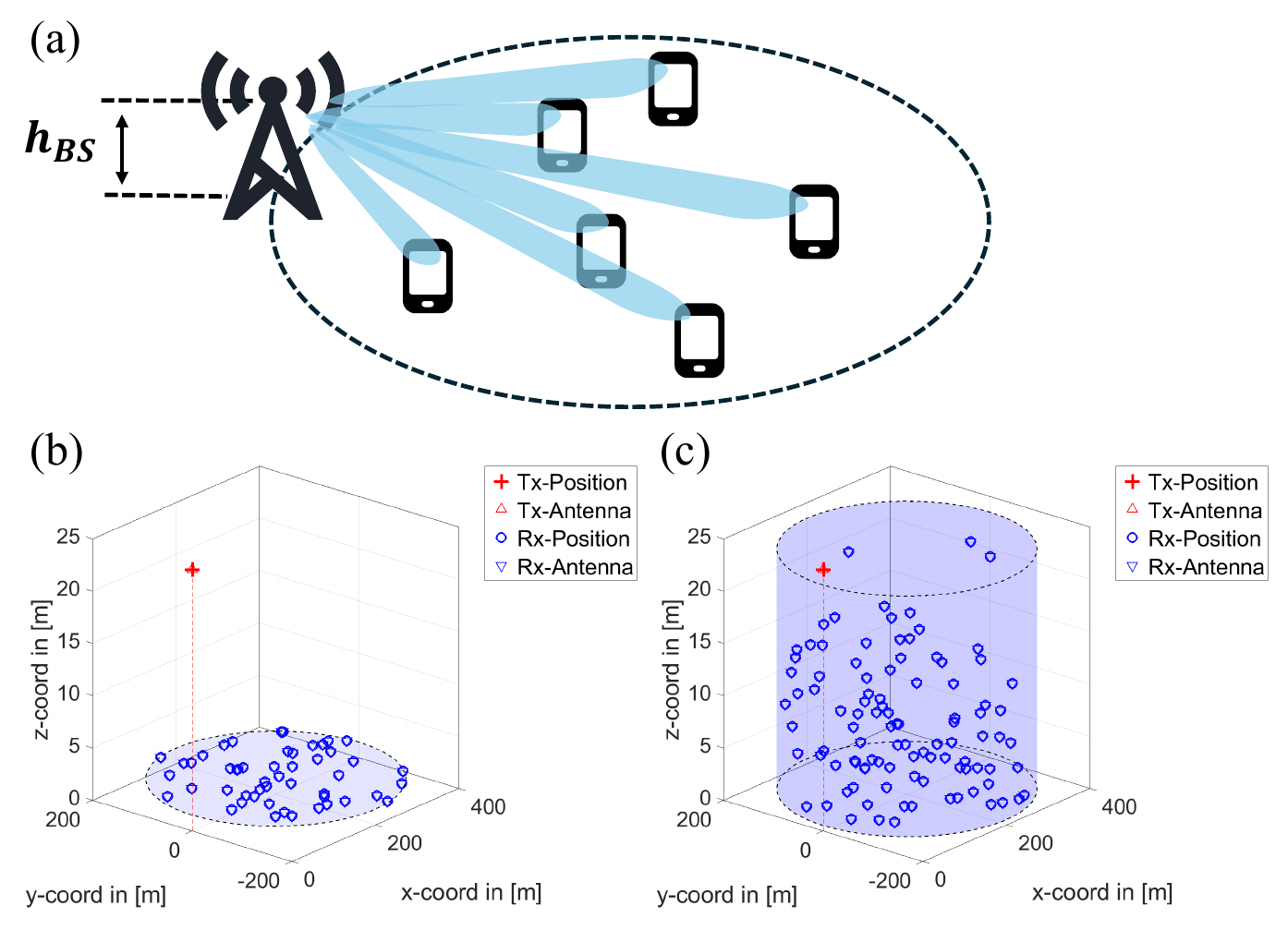}
\caption{3GPP TR 38.901 Urban Macro (UMa) scenario: (a) system setup showing base station height $h_{\mathrm{BS}} = 25~\mathrm{m}$, (b) 2-D user distribution, and (c) 3-D user distribution generated by QuaDRiGa.}
\label{3gpp_scenario}
\end{figure}

The 3GPP TR 38.901 Urban Macro (UMa) scenario is simulated using the QuaDRiGa platform~\cite{jaeckel2012geometric, jaeckel2014quadriga, jaeckel2017quasi} to obtain the channel matrix $\mathbf{H}$. Figure~\ref{3gpp_scenario}(a) illustrates the overall simulation setup, where a base station is placed at the cell edge with a height of 25~m and equipped with an antenna array of aperture $3\lambda \times 3\lambda$. The vertical element spacing is fixed at $0.5\lambda$, while the number of antennas varies along the horizontal axis. The system SNR is set to 20~dB. 
This SNR value is used to determine the noise variance $\sigma^2$ appearing in the SINR expression and the regularization factor in the MMSE precoding. In the simulations, 50 users equipped with omnidirectional antennas are randomly deployed within a circular cell of radius 200~m. 
For the 2-D deployment, all users are located on the horizontal plane with a fixed antenna height of 1.5~m, whereas for the 3-D deployment the user antenna heights are randomly drawn from a uniform distribution between 1.5~m and 22.5~m. All capacity results are averaged over 200 independent Monte Carlo snapshots, including random user drops and small-scale fading realizations.
Figures~\ref{3gpp_scenario}(b) and (c) show the corresponding layouts generated by QuaDRiGa, where users are located in a 2-D plane and in a 3-D volume, respectively.

\begin{figure}[ht!]
\centering
    \includegraphics[width=0.9\linewidth]{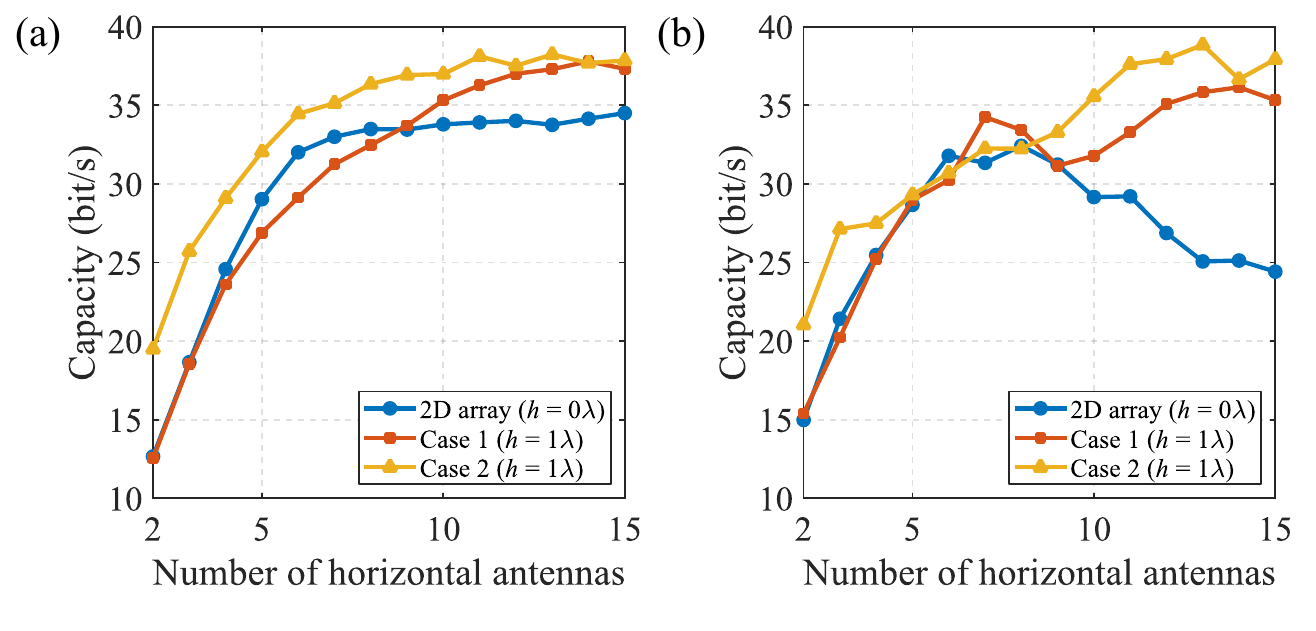}
\caption{Capacity performance of different base station antenna arrays under the 3GPP UMa scenario with users distributed in a 2-D plane: (a) with standard 3GPP antenna element, and (b) with dipole antenna element including coupling and efficiency.}
\label{3GPP_user2D}
\end{figure}

\begin{figure}[ht!]
\centering
    \includegraphics[width=0.9\linewidth]{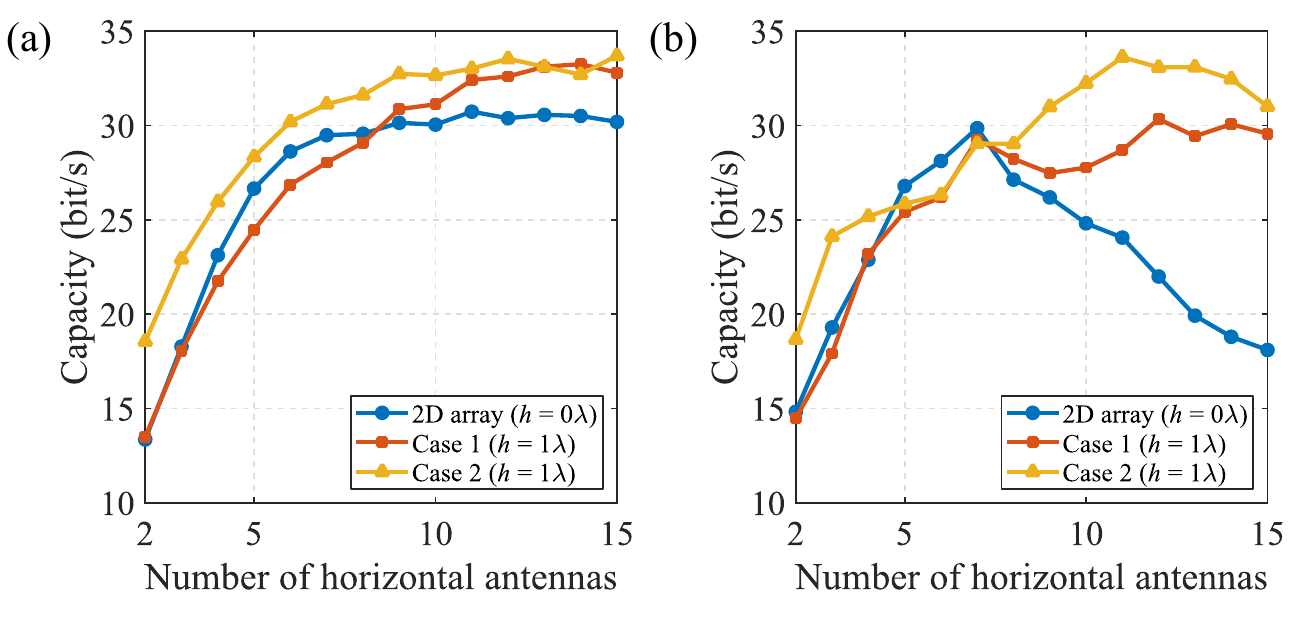}
\caption{Capacity performance of different base station antenna arrays under the 3GPP UMa scenario with users distributed in a 3-D volume: (a) with standard 3GPP antenna element, and (b) with dipole antenna element including coupling and efficiency.}
\label{3GPP_user3D}
\end{figure}

The channel capacity results under the 3GPP UMa scenario are shown in Figures~\ref{3GPP_user2D} and \ref{3GPP_user3D}. 
Figure~\ref{3GPP_user2D} presents the case where users are distributed in a 2-D plane. 
As observed in Figure~\ref{3GPP_user2D}(a), when using the standard 3GPP antenna element at the base station, all antenna elements are assumed to share identical embedded radiation patterns with unit radiation efficiency, consistent with the 3GPP reference antenna model. Under this assumption, mutual coupling and radiation efficiency degradation are not explicitly modeled.
At an inter-element spacing of $0.3\lambda$, Case~1 and Case~2 achieve capacity gains of 6.96\% and 12.41\% over the planar baseline, respectively. 
In contrast, Figure~\ref{3GPP_user2D}(b) shows the case with practical dipole antennas obtained from full-wave simulations, where both embedded element radiation patterns and embedded radiation efficiencies are explicitly incorporated into the channel model. It is observed that the capacity of the dipole-based array first increases and then decreases as the number of horizontally arranged antennas grows. This non-monotonic behavior reflects a tradeoff between increased spatial degrees of freedom and reduced radiation efficiency: while adding antennas initially improves multi-user spatial separation, further increasing the number of antenna elements within the same array aperture strengthens mutual coupling among closely spaced dipoles, which degrades the embedded radiation efficiency and reduces the effective channel gain.
In this case, both 3-D configurations outperform the planar array by more than 20\% at $0.3\lambda$ spacing, demonstrating the importance of efficiency in realistic array implementations.

Figure~\ref{3GPP_user3D} depicts the results when users are distributed in a 3-D volume. The absolute channel capacity decreases compared with the 2-D user distribution. Nevertheless, the overall trend remains consistent, and the 3-D base station arrays still provide clear capacity advantages. With an inter-element spacing of $0.3\lambda$, Case~1 and Case~2 achieve capacity gains of 7.36\% and 9.54\% over the planar baseline, respectively, confirming that the benefits of volumetric base station arrays are preserved even when users are distributed in three dimensions.

\section{Performance Analysis with Practical Antenna Designs}
\subsection{Antenna Design}
\begin{figure}[ht!]
\centering
    \includegraphics[width=0.9\linewidth]{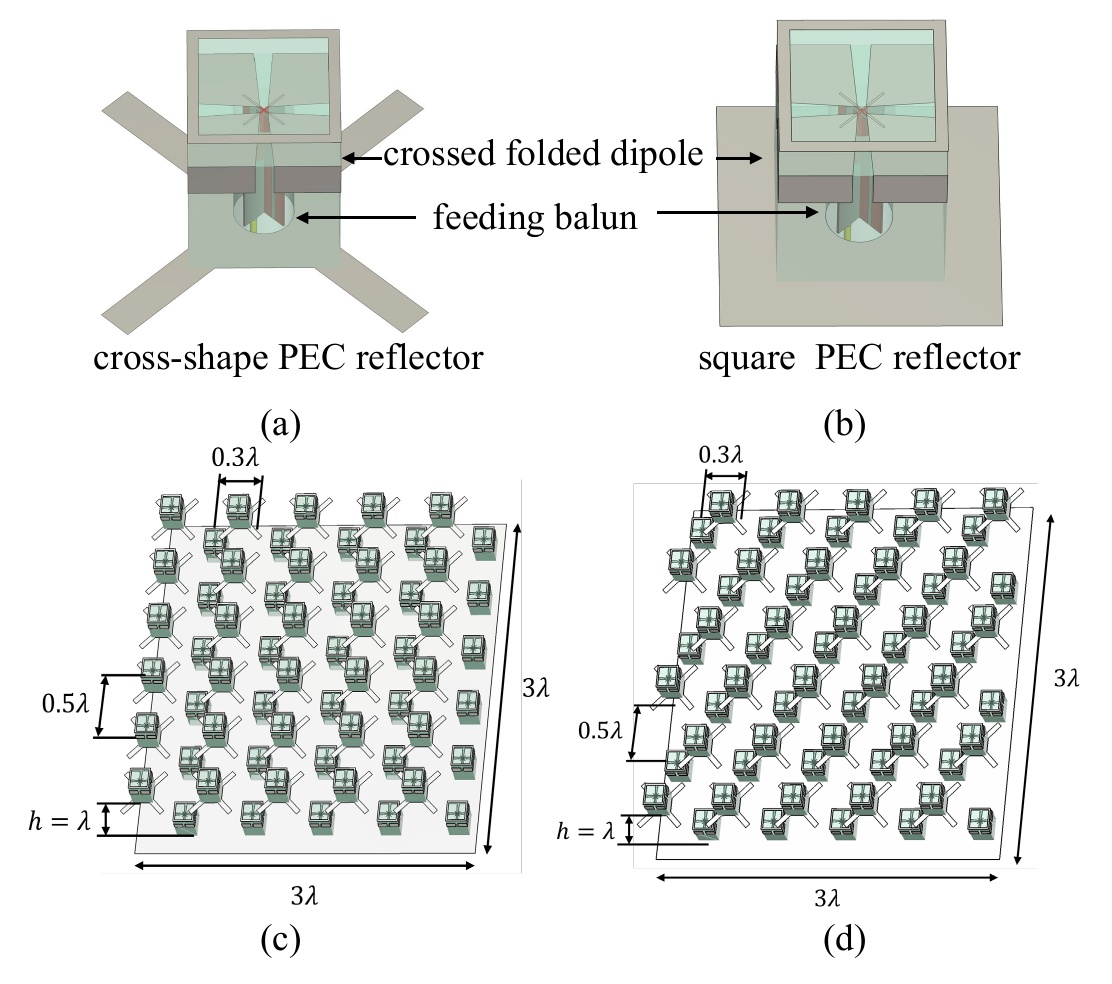}
\caption{(a) Upper-layer antenna with cross-shaped reflector. (b) Lower-layer antenna with square reflector. (c) Structural layout of Array~3.  (d) Structural layout of Array~4.}
\label{antenna_structure}
\end{figure}

The compact antenna element from~\cite{yang2022miniaturized} is adopted to model realistic deployment conditions. This element features a miniaturized structure with excellent radiation performance, making it suitable for practical implementation. The schematic of the antenna element is shown in Figure~\ref{antenna_structure}(a) and Figure~\ref{antenna_structure}(b). The main body consists of a cross-folded dipole structure with bottom-side coaxial feeding, while a balun is integrated at the center to ensure proper impedance matching between the feeding line and the radiating element. A square reflector is mounted beneath the lower-layer antennas to steer radiation toward the upper hemisphere, while a cross-shaped base is employed for the upper-layer antennas to reduce blockage of the lower-layer elements.

We consider four array configurations, denoted as Array~1–Array~4, which share a common physical aperture of $3\lambda\times3\lambda$. Their definitions are given below, while the structural layouts of Array~3 and Array~4 are illustrated in Figure~\ref{antenna_structure}(c) and (d), respectively.

\begin{itemize}
  \item \textbf{Array~1}: a 2-D planar reference array with element spacing $0.5\lambda$ in both the $x$ and $y$ directions, consisting of 36 antennas in total.
  \item \textbf{Array~2}: a 2-D planar tightly-spaced array whose element spacing is $0.3\lambda$ in the $x$ direction and $0.5\lambda$ in the $y$ direction, consisting of 60 antennas in total.
  \item \textbf{Array~3}: a 3-D dual-layer array with Case~1 configuration, whose element spacing is $0.3\lambda$ in the $x$ direction and $0.5\lambda$ in the $y$ direction; the inter-layer height offset is $h=1\lambda$, consisting of 60 antennas in total.
  \item \textbf{Array~4}: a 3-D dual-layer array with Case~2 configuration, whose element spacing is $0.3\lambda$ in the $x$ direction and $0.5\lambda$ in the $y$ direction; the inter-layer height offset is $h=1\lambda$, consisting of 60 antennas in total.
\end{itemize}

\begin{figure}[ht!]
\centering
    \includegraphics[width=0.8\linewidth]{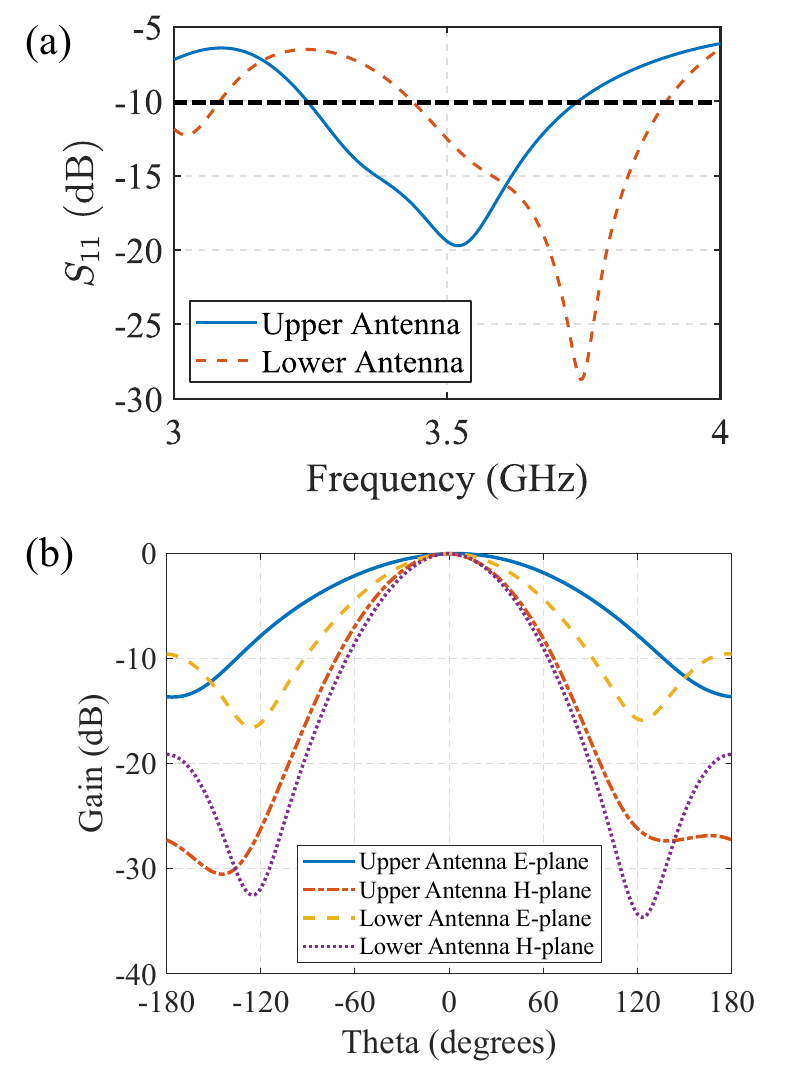}
\caption{(a) Simulated reflection coefficients of the isolated lower and upper antennas. (b) Gain of the isolated lower and upper antennas at far field (reference distance is taken as 1 m).}
\label{Sparameter_gain_single}
\end{figure}

\subsection{Radiation Performance}
\subsubsection{Isolated Antenna Performance}
The reflection coefficients and far-field radiation patterns of the isolated upper and lower antennas are shown in Figure~\ref{Sparameter_gain_single}. Figure~\ref{Sparameter_gain_single}(a) shows that the upper antenna with the cross-shaped reflector resonates near the target frequency of 3.5\,GHz with a $-10$\,dB impedance bandwidth exceeding 500\,MHz. In contrast, the lower antenna with the square reflector shows a resonance shift toward 3.75\,GHz, while still maintaining a comparable $-10$\,dB bandwidth of nearly 500\,MHz. At the target frequency of 3.5\,GHz, both antennas achieve reflection coefficients below $-10$\,dB, confirming good impedance matching within the desired operating band. The corresponding far-field gain patterns, depicted in Figure~\ref{Sparameter_gain_single}(b), further demonstrate favorable radiation characteristics, with the upper antenna exhibiting more uniform and stable directional behavior.  

\subsubsection{Array Performance}
\begin{figure}[ht!]
\centering
    \includegraphics[width=0.9\linewidth]{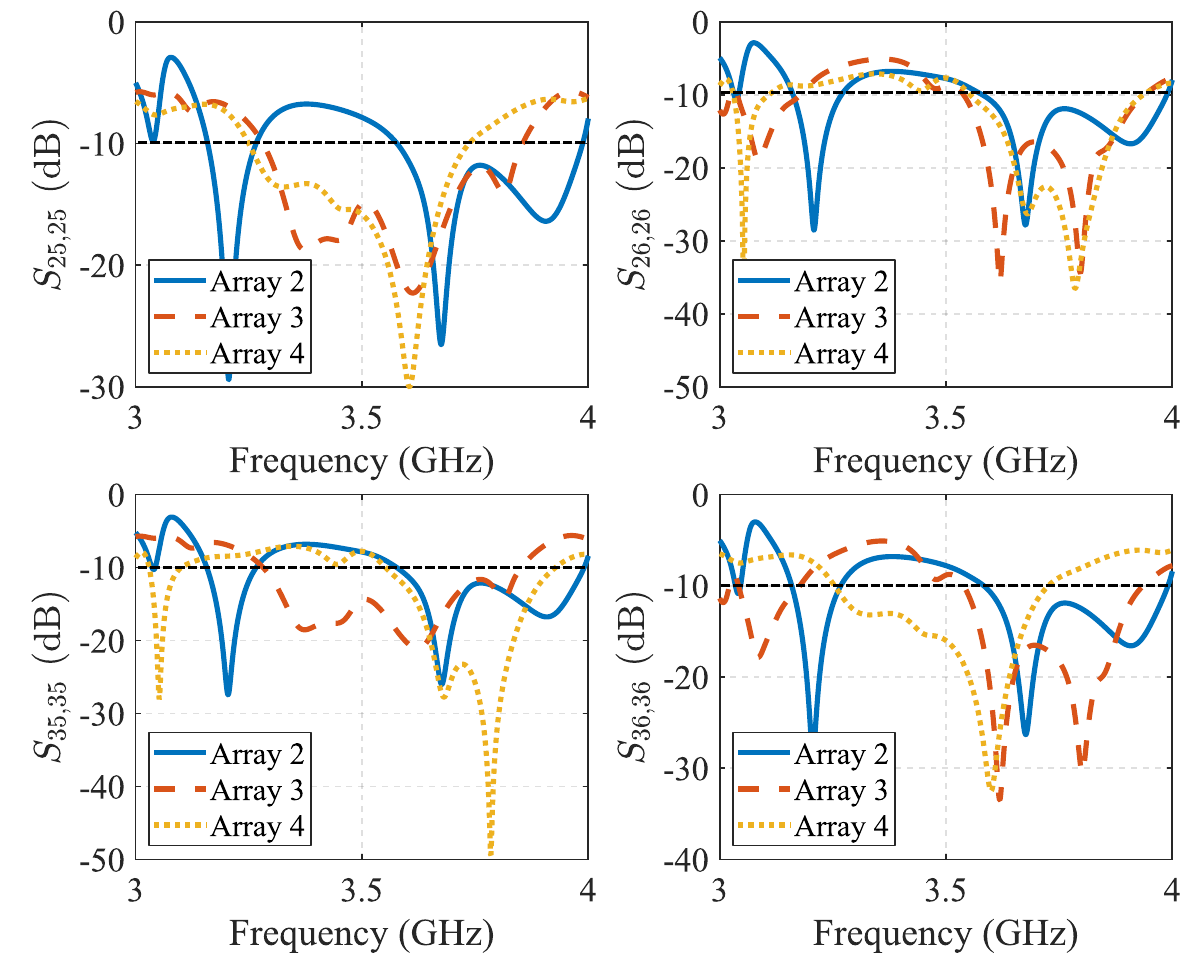}
\caption{Reflection coefficients of the four central antennas in Array~2, Array~3, and Array~4.}
\label{Sparameter_inarray}
\end{figure}

\begin{figure}[ht!]
\centering
    \includegraphics[width=0.9\linewidth]{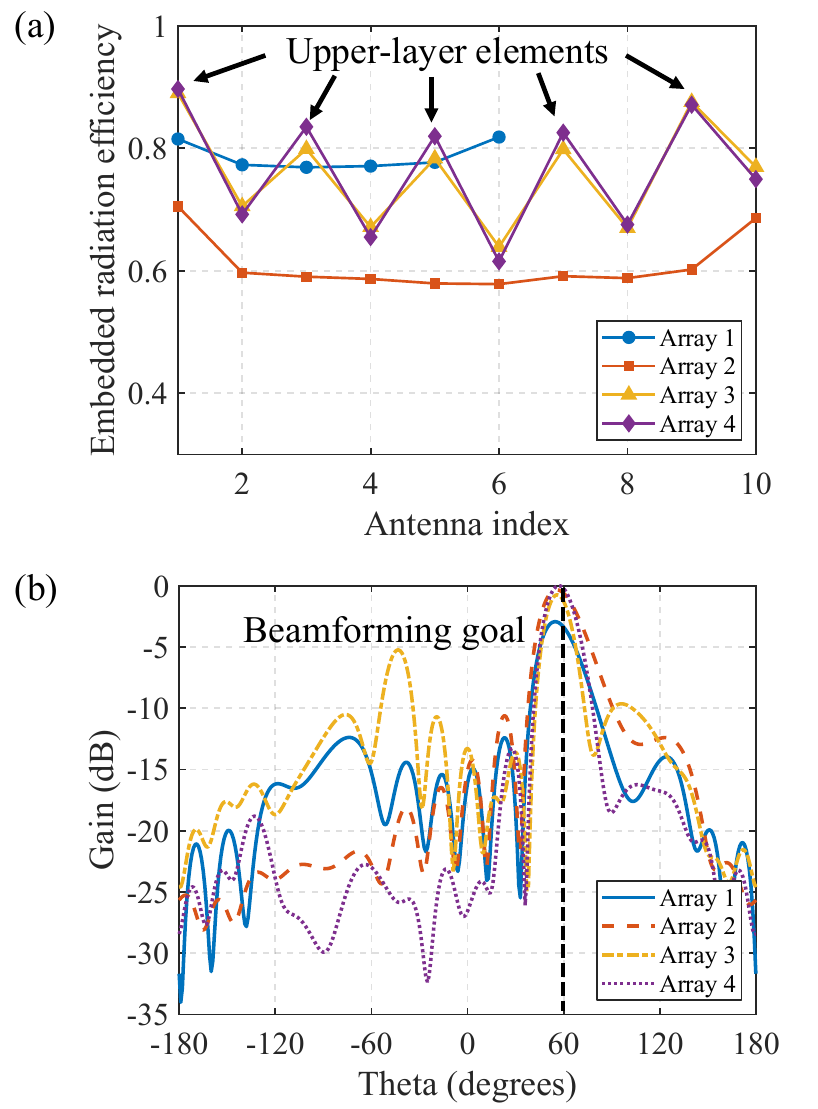}
\caption{(a) Embedded radiation efficiency of center-row antennas for different array configurations. (b) Beamforming performance with target direction $\theta = 60^\circ, \phi = 0^\circ$.}
\label{effic_beamforming}
\end{figure}

The reflection coefficients of the four central antennas in Array~2, Array~3, and Array~4 are shown in Figure~\ref{Sparameter_inarray}. Around the target frequency of 3.5\,GHz, all configurations achieve acceptable matching, but the 3-D layouts (Arrays~3 and~4) provide more stable and deeper notches than the dense 2-D array. Furthermore, the embedded radiation efficiency of the center-row antennas in the four configurations is shown in Figure~\ref{effic_beamforming}(a). The planar half-wavelength array achieves an efficiency of approximately 0.8. Reducing the horizontal spacing to $0.3\lambda$ results in stronger mutual coupling, which drops the efficiency to around 0.6. By elevating part of the antennas by one wavelength, the upper-layer antennas exhibit significantly improved efficiency, while the increased inter-element distance also reduces coupling for the lower-layer antennas, resulting in a slight efficiency gain. The alternating high/low efficiency values observed for the 3-D arrays arise from the dual-layer staggered geometry, where the center row consists of antennas from the upper and lower layers arranged in alternating positions. Overall, the average efficiency approaches that of the planar half-wavelength array.

The beamforming characteristics of the four considered arrays are illustrated in Figure~\ref{effic_beamforming}(b), with the beam steered toward the target direction of $\theta = 60^\circ, \phi = 0^\circ$. For the planar half-wavelength array, the 3-dB beamwidth of the main lobe is approximately $26.4^\circ$, which slightly broadens to $27.6^\circ$ in the planar $0.3\lambda$ array owing to the stronger mutual coupling effects. By contrast, the 3-D Case~1 and Case~2 arrays produce significantly narrower beams, with 3-dB beamwidths of $18.2^\circ$ and $20.9^\circ$, respectively, thereby enhancing the effective spatial degrees of freedom and improving the achievable channel capacity. In addition, the radiation pattern comparison shows that Case~1 exhibits relatively higher sidelobes, while Case~2 concentrates more energy within the main lobe. This result agrees well with the preceding theoretical analysis, where Case~2 demonstrates a slightly higher channel capacity than Case~1.

To further evaluate the beamforming performance under different steering conditions, Table~\ref{tab1} lists the 3-dB beamwidths when the main lobe is directed toward $\theta=0^\circ$, $30^\circ$, $45^\circ$, and $60^\circ$ (with $\phi=0^\circ$). The results show that the two planar arrays (Array~1 and Array~2) maintain comparable beamwidths across all scan angles. In contrast, the 3-D dual-layer arrays (Array~3 and Array~4) consistently achieve narrower beams, and the performance gap widens as the steering angle increases. This demonstrates the superior spatial resolution of the 3-D array architecture, particularly in wide-angle scanning scenarios.

\begin{table}[ht!]
    \centering
    \caption{3-dB beamwidths of the four array configurations under different beamforming angles.}
    \label{tab1}
    \resizebox{\linewidth}{!}{%
    \begin{tabular}{|c|c|c|c|c|}
        \hline
        \multirow{2}{*}{Beamforming goal ($^\circ$)} & 
        \multicolumn{4}{c|}{3-dB beamwidth ($^\circ$)} \\ \cline{2-5}
         & Array~1 & Array~2 & Array~3 & Array~4 \\
        \hline
        $\theta=0, \phi=0$ & 17.1 & 16.9 & 16.6 & 16.6\\
        \hline
        $\theta=30, \phi=0$ & 19.4 & 19.2 & 16.7 & 17.8\\
        \hline
        $\theta=45, \phi=0$ & 23.1 & 22.2 & 18.9 & 19.5\\
        \hline
        $\theta=60, \phi=0$ & 26.4 & 27.6 & 18.2 & 20.9\\
        \hline
    \end{tabular}%
    }
\end{table}

\subsection{MIMO Performance}
\subsubsection{Kronecker Model}
\begin{figure}[ht!]
\centering
    \includegraphics[width=0.9\linewidth]{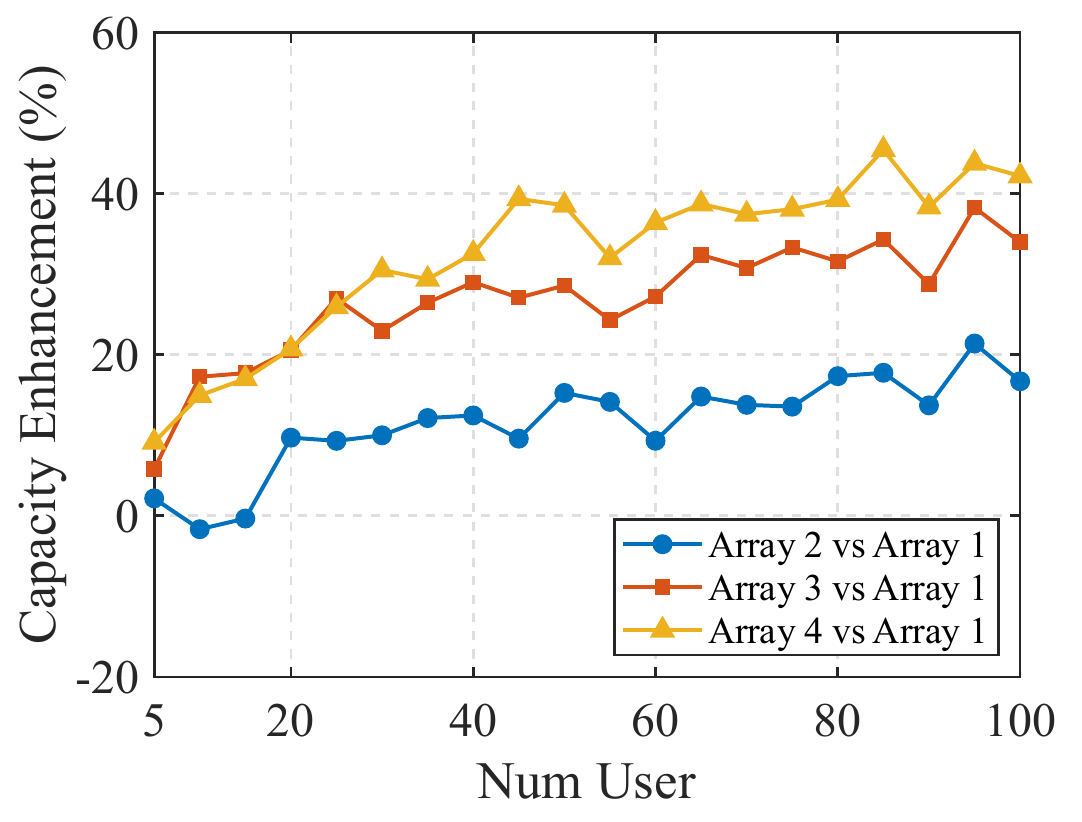}
\caption{Capacity enhancement of Array~2, Array~3, and Array~4 (each with 60 antennas) relative to Array~1 (36 antennas) in the Kronecker model, as the number of users increases from 5 to 100.}
\label{kronecker_cap_enhancement}
\end{figure}

The four array configurations are first evaluated in the Kronecker model, as illustrated in Figure~\ref{kronecker_cap_enhancement}. In this model, a rich-scattering Rayleigh fading environment is assumed, where only small-scale fading due to multipath is considered and large-scale fading effects are neglected. Under such conditions, the loss of radiation efficiency caused by mutual coupling can be compensated by the increased spatial degrees of freedom when the antenna spacing is reduced. Consequently, the capacity of the $0.3\lambda$ planar array (Array~2) surpasses that of the half-wavelength planar array (Array~1), while the two 3-D arrays (Arrays~3 and~4) demonstrate even more substantial capacity improvements. Moreover, the relative gain grows with the number of users: at 60 users, Arrays~2,~3, and~4 achieve capacity enhancements of 9.32\%, 27.24\%, and 36.42\%, respectively, compared with Array~1. When the user count reaches 100, the improvements further increase to 16.71\%, 34.00\%, and 42.21\%.

\subsubsection{3GPP Model}
\begin{figure}[ht!]
\centering
    \includegraphics[width=0.9\linewidth]{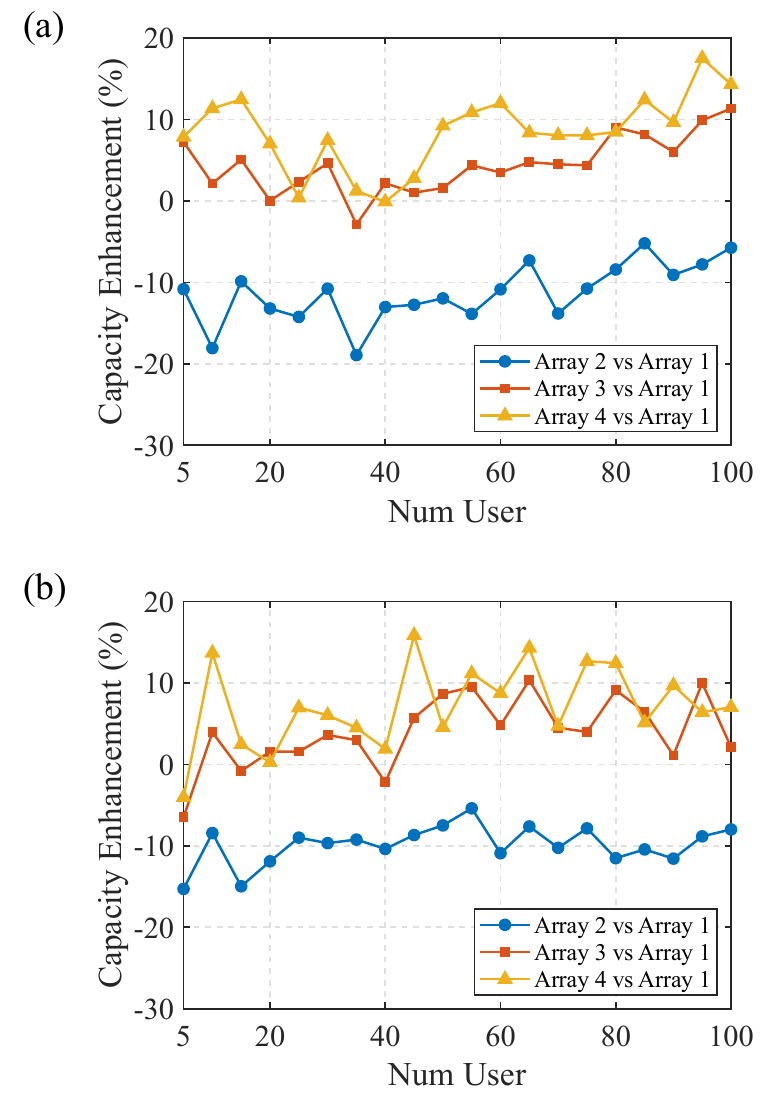}
\caption{Capacity enhancement of Array~2, Array~3, and Array~4 (each with 60 antennas) relative to Array~1 (36 antennas) in the 3GPP UMa scenario, with the number of users increasing from 5 to 100: (a) users distributed in a 2-D plane, and (b) users distributed in a 3-D volume.}
\label{3gpp_cap_enhancement}
\end{figure}

The four array configurations are further assessed in the 3GPP UMa scenario, as shown in Fig.~\ref{3gpp_cap_enhancement}. Unlike the Kronecker model, the 3GPP channel incorporates large-scale fading effects, making radiation efficiency and beam directivity more critical to system performance. 

Figure~\ref{3gpp_cap_enhancement}(a) illustrates the capacity enhancement when users are randomly distributed in a 2-D plane. In this case, Array~2 suffers from efficiency degradation, resulting in lower capacity than Array~1. By contrast, the two 3-D arrays (each with 60 antennas) deliver superior performance, owing to improved efficiency and narrower beamwidths. At 60 users, Arrays~3 and~4 achieve capacity gains of 3.49\% and 12.00\% over Array~1, and 16.10\% and 25.65\% over Array~2. When the number of users increases to 100, the gains further rise to 11.34\% and 14.31\% relative to Array~1, and 18.12\% and 21.28\% relative to Array~2.  

Figure~\ref{3gpp_cap_enhancement}(b) shows the case where users are distributed in a 3-D volume. Here, the relative improvements of the 3-D arrays are smaller but remain consistent. At 60 users, Arrays~3 and~4 provide capacity gains of 4.80\% and 8.74\% over Array~1, and 17.63\% and 10.97\% over Array~2. With 100 users, the gains become 2.10\% and 7.20\% relative to Array~1, and 22.06\% and 16.31\% relative to Array~2.  

Overall, these results confirm that 3-D array architectures can sustain performance advantages under tightly coupled conditions, even in realistic propagation scenarios. While the present designs already demonstrate clear benefits in terms of efficiency and capacity, further improvements remain possible. As observed in the beamforming analysis, the 3-D configurations—particularly Case 1—exhibit elevated sidelobe levels that may reduce spatial selectivity in multiuser scenarios. Integrating 3-D arrays with advanced techniques such as electromagnetic hybrid beamforming~\cite{ji2024electromagnetic} could address this limitation by enabling amplitude-phase co-optimization across the aperture, thereby suppressing sidelobes while more effectively exploiting the available spatial modes. Such integration may also provide additional mitigation of coupling-induced efficiency loss, thereby further enhancing the overall performance of 3-D holographic MIMO systems.

From a deployment perspective, it is worth noting that large-scale 3-D arrays may introduce additional hardware complexity compared with planar implementations, including more intricate feeding networks, calibration requirements, and potential increases in RF-chain cost, particularly at high-frequency bands. The present work focuses on establishing the electromagnetic and channel-level performance potential under fixed-aperture constraints, while practical trade-offs among performance, hardware complexity, and cost remain important topics for future investigation at the prototype and system levels.

\section{Conclusion}
This paper has presented a systematic evaluation of 3-D holographic MIMO arrays under Clarke, Kronecker, and 3GPP channel models, offering a unified perspective from idealized theory to realistic deployment. The results demonstrate that volumetric array configurations significantly enlarge the effective degrees of freedom and mitigate the degradation caused by mutual coupling when the aperture is fixed. Compared with planar baselines, 3-D arrays achieve narrower beamwidths, higher radiation efficiency, and consistent capacity improvements across diverse propagation environments. In 3GPP urban macro scenarios, capacity gains of nearly 20\% are achieved by the proposed 3-D arrays compared to conventional 2-D planar arrays under the same element spacing and aperture size, confirming the robustness and scalability of volumetric designs. These findings bridge electromagnetic modeling and system-level evaluation, providing design guidelines for next-generation base station arrays in 6G systems.



\end{document}